\documentclass{v16nufact}

\def\Journal#1#2#3#4{{#1} {\bf #2}, #3 (#4)}

\def\be{\begin{equation}}
\def\ee{\end{equation}}
\def\bea{\begin{eqnarray}}
\def\eea{\end{eqnarray}}

\newcommand{\Photo}{}

\begin{document}
\vspace*{4cm}
\title{LArIAT: World’s First Pion-Argon Cross-Section}

\author{P. A. Hamilton}

\address{Fermilab, Wilson Hall 10W, Batavia, Illinois 60510, USA}

\maketitle\abstracts{
The LArIAT experiment has performed the world's first measurement of the total charged-current pion cross-section on an argon target, using the repurposed ArgoNeuT detector in the Fermilab test beam. Presented here are the results of that measurement, 
along with an overview of the LArIAT experiment and details of the LArIAT collaboration's plans for future measurements.}

\section{Introduction}

The U.S. neutrino programme revolves around the use of liquid argon TPC technology. TPCs of varying scales are planned (or already in operation) for the Fermilab short baseline programme, while the far detector for the proposed long-baseline experiment DUNE would be the largest liquid argon TPC ever constructed~\cite{dune}. Using these detectors to their fullest effect will require a more detailed understanding of the interactions of neutrinos on argon nuclei. One way to contribute to our understanding of 
neutrino-nucleus interactions in argon is to study the interactions of charged particles on argon: these are the interactions occurring in the intranuclear environment during neutrino-nucleus interactions, and probe the same nuclear structure. Their interaction 
rate is also much higher, avoiding the common problem in neutrino cross-section experiments of being statistics-limited.

The LArIAT experiment observes the interactions of a tertiary beam composed mostly of pions (but also kaons, protons, muons and electrons) on liquid argon. By measuring these interactions, LArIAT can provide data to help tune hadron-nucleus interaction models in Geant4 
and neutrino generators, while also performing important research and development for reconstruction and calorimetry in liquid argon TPCs.  The details of LArIAT's first measurement, the total charged-current pion-argon cross-section, are explained below; measurements 
of other incident particles (e.g. kaons) and more specific interaction channels are expected in the near future.

\section{The LArIAT Experiment}

The LArIAT experiment uses the ArgoNeuT liquid argon TPC, placed in a test beam at the Fermilab Test Beam Facility (FTBF). The TPC is complemented with a suite of auxiliary detectors, used to measure the properties of the beam before it enters the TPC (and also the properties of those particles which pass through the TPC).
Each of these systems are detailed in the sections below.

\subsection{The Fermilab Test Beam}

\begin{figure}[h!]
 \centerline{\includegraphics[width=0.4\textwidth]{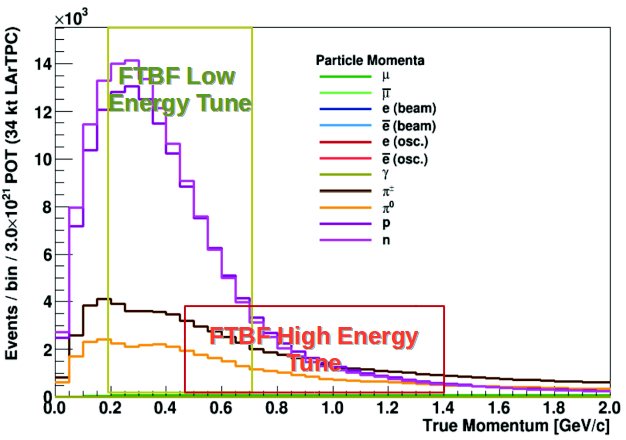}}
 \caption{Momentum spectra (histograms) for secondary particles typical of those in the Booster neutrino experiments ICARUS, SBND and MicroBooNE. The superimposed boxes show the the range of particle momenta provided by the FTBF tertiary spectrometer in the low-energy tune (which is well matched to these spectra) and the high-energy tune (which is typical of particle momenta produced by neutrinos in the off-axis NuMI beam for NOvA, and those expected in the DUNE far detector).}
 \label{fig:ftbfspectrum}
\end{figure}

The Fermilab test beam is a tertiary beam created from particle interactions on two successive targets. A primary beam of 120 GeV protons impinges onto an aluminium target to create a secondary beam composed primarily of pions; the energy of the secondary beam 
can be magnetically tuned between 8 and 80 GeV. This beam then impinges on a copper target to create the tertiary beam seen by LArIAT, which is correspondingly tunable in the range 0.2--1.5 GeV. The full spectrum of the test beam is shown in Figure \ref{fig:ftbfspectrum}.

\subsection{Auxiliary Beamline Detectors}
\begin{figure}[h!]
 \centerline{\includegraphics[width=0.6\textwidth]{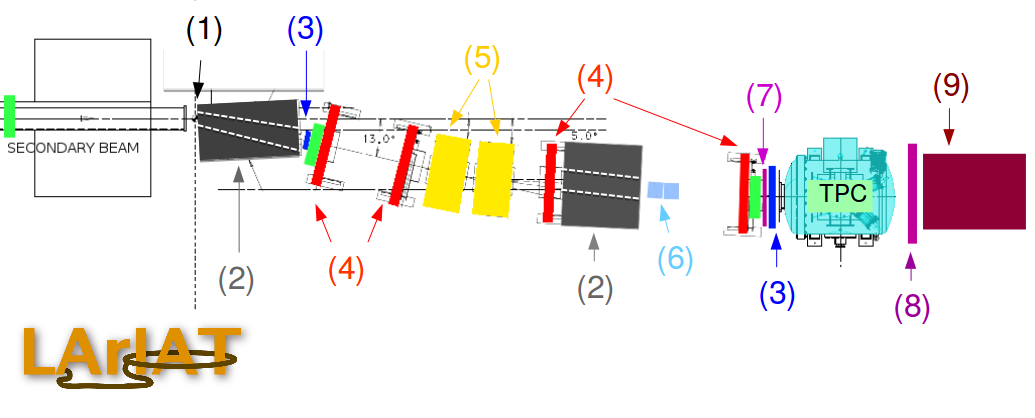}}
 \caption{The LArIAT beamline, with elements numbered as follows: (1) copper target, (2) collimators, (3) time of flight detectors, (4) multi-wire proportional chambers, (5) bending magnets, (6) aerogel Cherenkov detectors, (7) halo veto, (8) muon punch-through veto, and (9) the muon range stack.}
 \label{fig:beamline}
\end{figure}

The full suite of beamline detectors used by LArIAT to parameterise the test beam are shown in Figure \ref{fig:beamline}.

The auxiliary detectors upstream of the TPC are used to measure the momentum of incoming particles, and to perform particle identification. The detectors downstream are used to discriminate between through-going pions and muons. Further details of the most important systems are given below.

\subsubsection{Magnets and Multi-Wire Proportional Chambers}

The beamline bending magnets serve several purposes: sign-selecting and tuning the energy of particles reaching the TPC, while also measuring the momentum of incoming particles through their deflection in the magnetic field. This deflection is measured by pairs of multi-wire proportional chambers 
upstream and downstream of the magnets. Hits in these wire chambers define two straight lines between which an angle of deflection can be calculated. The distribution of reconstructed momenta from this calculation is shown in Figure \ref{fig:beambits}.

\subsubsection{Time of Flight Detectors}
\begin{figure}[h!]
 \centerline{\includegraphics[height=0.2\textheight]{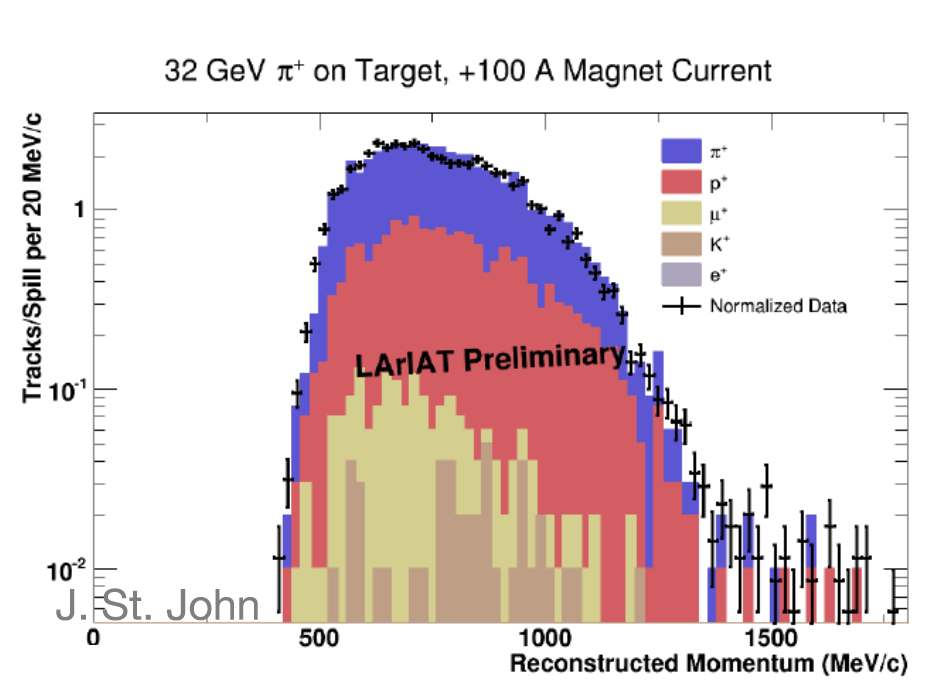}
 \includegraphics[height=0.17\textheight]{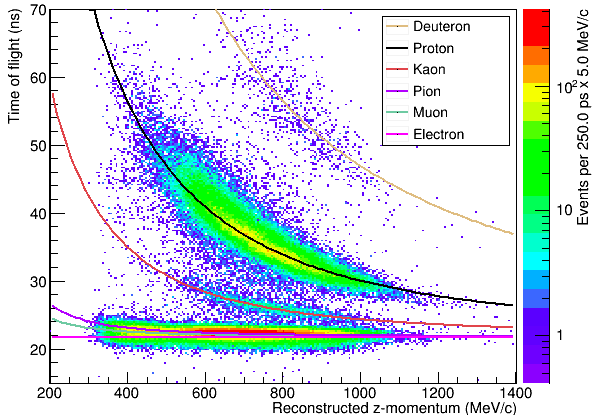}}
 \caption{Left, the reconstructed momentum profile of the beam in the high energy tune, showing a good agreement between simulation and data. Right, a histogram of particle time of flight against reconstructed z-momentum (Z being the beam direction), showing the clearly separated populations of muons/pions/electrons, kaons, and protons. 
The right-hand histogram includes events from both the high and low energy tunes.}
 \label{fig:beambits}
\end{figure}

Time of flight detectors positioned at the start and end of the beamline provide discrimination between particle species through measurements of the time taken to travel between target and TPC (``time of flight''). This provides clear discrimination between protons, kaons, and the combined category 
of muons, pions and electrons, as shown in Figure \ref{fig:beambits}.

\subsubsection{Aerogel Cherenkov Detectors}
Two aerogel-based Cherenkov detectors positioned directly behind the second collimator provide a second method of particle identification, aimed at discriminating between muons and pions. The two detectors have differing indices of refraction ($n = 1.11$ and 1.057), 
such that:
\begin{itemize}
 \item In the momentum range 200-300 MeV/c, muons will produce Cherenkov radiation only in the first detector, while pions will produce Cherenkov radiation in neither.\vspace{-2mm}
 \item In the momentum range 300-400 MeV/c, muons will produce Cherenkov radiation in both detectors, while pions will produce Cherenkov radiation only in the first.
\end{itemize}
Combining observations from the aerogel with momentum measurements from the wire chambers therefore allows for the separation of muons and pions in these energy ranges. However, this technique has not yet been implemented in a LArIAT analysis.

\subsubsection{Muon Range Stack}
The muon range stack is a layered assembly of steel blocks and planes of scintillator paddles, read out in sixteen channels across four layers. This provides simple discrimination between through-going muons and pions by their penetration depth in the steel. 
Like the aerogel detectors, this system is taking data but not currently used in LArIAT analyses.

\subsection{The LArIAT TPC}

LArIAT uses the repurposed ArgoNeuT TPC, as mentioned previously. This is a 40 cm $\times$ 47 cm $\times$ 90 cm liquid argon TPC, with an active volume of 170 L of argon under a 500 V/cm drift field. The TPC is read out with two wire planes oriented at $\pm 60^{\circ}$, each with a 4 mm wire pitch. This TPC is described in detail in Anderson {\it et al.}, 2012~\cite{agnt}.

\begin{figure}[h!]
\centerline{\includegraphics[height=0.2\textheight]{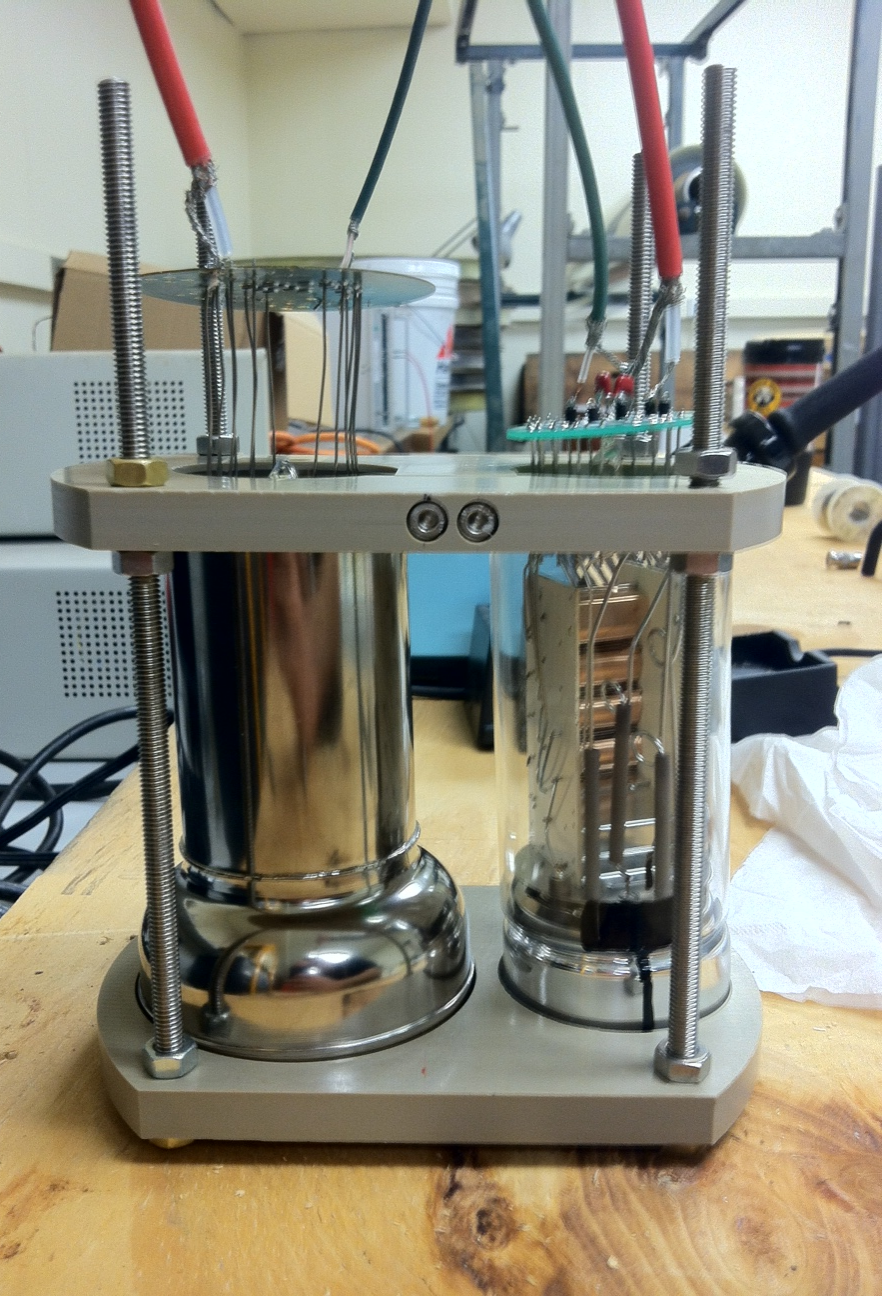}
\includegraphics[height=0.2\textheight]{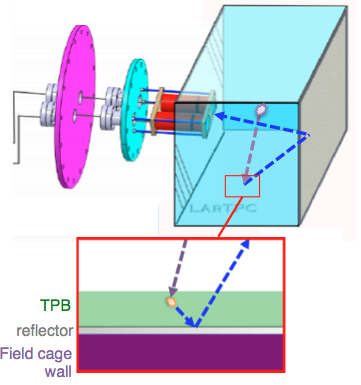}
\includegraphics[height=0.2\textheight]{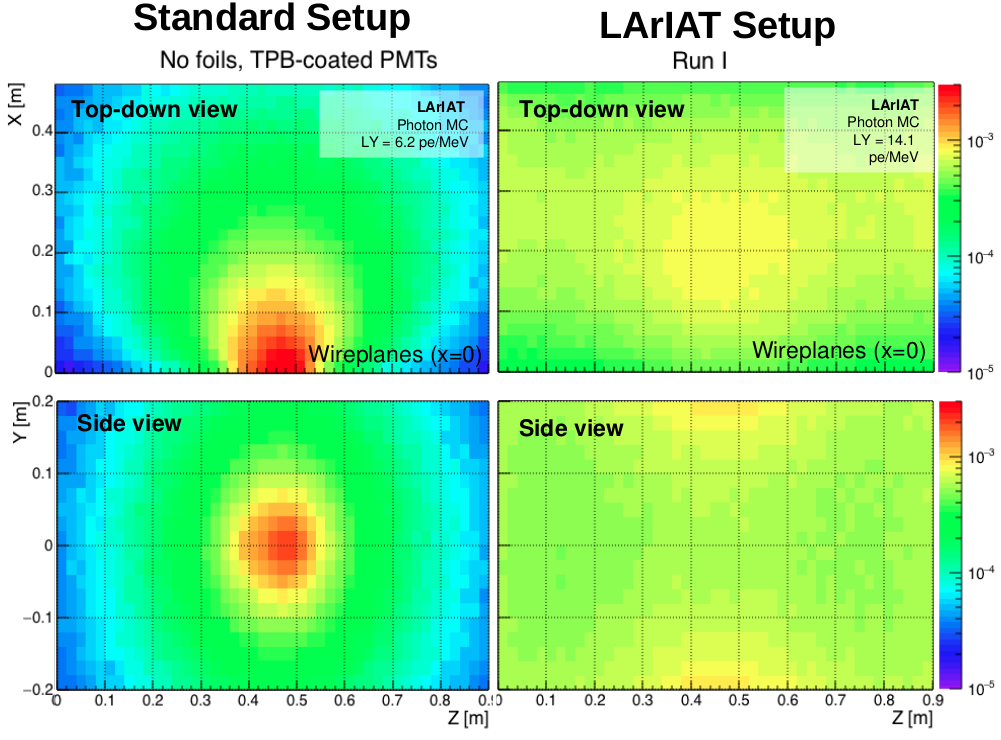}}
 \caption{A photograph of the LArIAT PMTs (left), a schematic of the LArIAT PMT/reflector setup (centre), and plots (right) contrasting the performance of the LAriAT setup (right-hand column) to a ``conventional'' setup such as the one used in MicroBooNe (left-hand column) as simulated in the LArIAT TPC. The colour scale shows the fractional proton visibility, while the Y and Z axes spatially define the readout 
 plane of the TPC (with positive Z being the beam direction). The LArIAT setup clearly provides much more uniform coverage than a conventional PMT array.}
 \label{fig:reflector}
\end{figure}

One feature new to LArIAT is the light collection system. ``Conventional'' liquid argon TPC designs feature an array of PMTs spaced to give direct coverage of the full active volume, with TPB-coated plates in front of the PMT windows. In contrast uses only two PMTs --- a 3'' Hamamatsu R-11065 and a 2'' ETL D757KFL --- 
placed close together and with no TPB-coated plates. Instead, the walls of the field cage are covered with a layer of TPB-coated reflector foil, allowing scintillation light produced by particle tracks to bounce around the light-tight interior of the TPC with minimal losses until it arrives at the PMTs, with its wavelength 
pre-shifted into the range the PMTs can detect. This setup is shown in Figure \ref{fig:reflector}, along with the improvement in visibility over the detector volume it provides.

\section*{Cross-Section Measurement}

\subsection{Definition}
\label{subsec:def}

Considering a particle impinging on a slab of target material, the probability of the particle's passing through the slab without interaction ($P_{\textrm{surviving}}$) can be expressed
\begin{equation}
P_{\textrm{surviving}} = e^{-\sigma n z} = 1 - P_{\textrm{interacting}}\textrm{,}
\end{equation}
where $\sigma$ is the cross-section per nucleus, n is the number density of nuclei, z is the depth of the slab, and $P_{\textrm{interacting}}$ the probability of an interaction in the slab.

$P_{\textrm{interacting}}$ can be estimated as the ratio of the number of particles interacting in the slab ($N_\textrm{interacting}$) to the number of particles impinging on the slab ($N_{\textrm{incident}}$):
\begin{equation}
\frac{N_{\textrm{interacting}}}{N_{\textrm{incident}}} = P_{\textrm{interacting}} = 1 - e^{-\sigma n z}{\textrm{.}}
\end{equation}

In the case of a thin slab, where $z$ is small and the energy $E$ of the particle entering the slab is approximately equal to its energy exiting the slab, this relation can be expanded to give the following expression for the cross-section:

\begin{equation}
\sigma(E) \approx \frac{1}{nz}P_{\textrm{interacting}} = \frac{1}{nz}\frac{N_{\textrm{interacting}}}{N_{\textrm{incident}}}\textrm{.}
\label{eq:xsec}
\end{equation}

\begin{figure}[h!]
\centerline{\includegraphics[height=0.15\textheight]{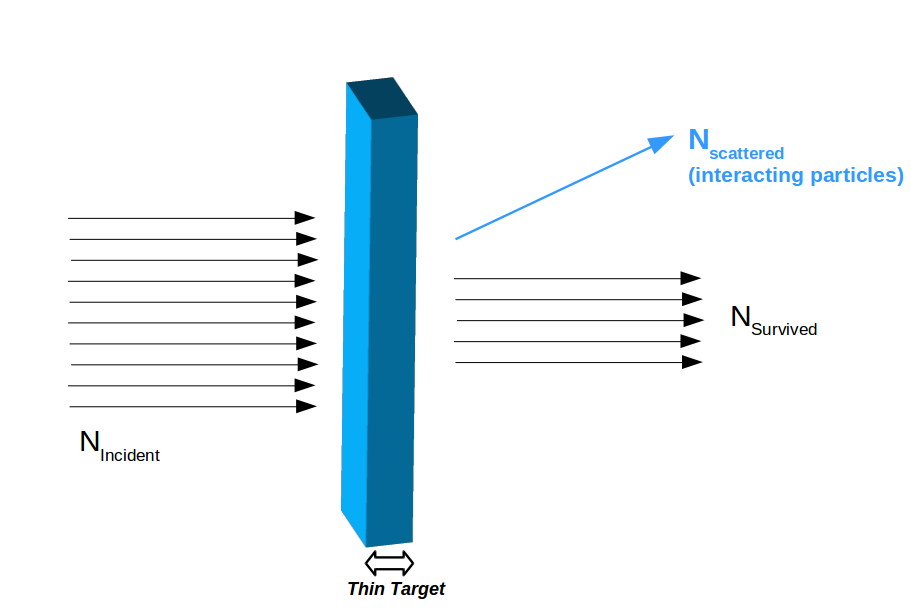}
\includegraphics[height=0.15\textheight]{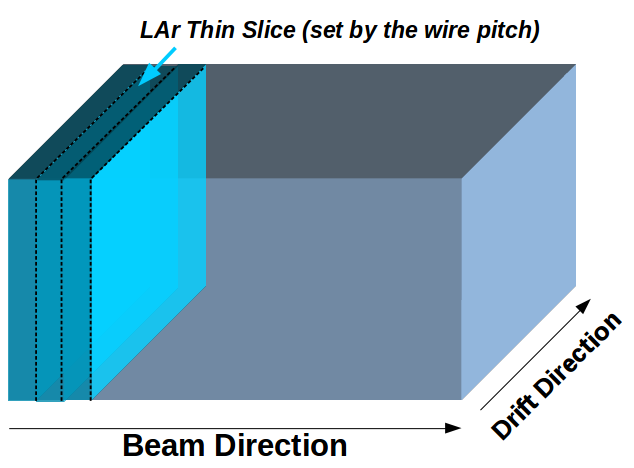}}
\caption{A cartoon of the ``thin slab'' approach used by LArIAT to assess the pion cross-section.}
\label{fig:sliceanddice}
\end{figure}

To measure this in data, we treat the LArIAT TPC as a series of ``thin slabs'' (as shown in Figure \ref{fig:sliceanddice}), with the slab depth equivalent to the horizontal distance between wires ($\sim 5$ mm). We define histograms of the number of slabs in which an incident pion is observed (i.e. the pion track is present), and the 
number of slabs in which an incident pion interacts (i.e. the pion track terminates, in a vertex or otherwise). These histograms are binned in kinetic energy, with the energy in the $n$th slab traversed by the pion being calculated as follows:

\begin{equation}
 E_n = (\sqrt{p_{\textrm{reco}}^2 - m_{\pi}^2} - m_{\pi}) - E_{\textrm{flat}} - \sum^{n-1}_{i = 0}\Big(\frac{dE}{dX}\Big)_i\times z_i \textrm{.}
 \label{eq:ke}
\end{equation}

Here $p_{\textrm{reco}}$ is the reconstructed momentum obtained from the wire chambers, $E_{\textrm{flat}}$ is a constant adjustment for energy loss in the dead material directly upstream of the TPC (e.g. the front face of the cryostat), and $z_{i}$ is the slab depth. 

By dividing the interacting histogram by the incident histogram, we recover $\sigma(E)$ as defined in Equation \ref{eq:xsec}. To apply this method, we must first select pion tracks, and be confident in our assessment of which slab contains the point of interaction takes place. This is achieved by the selection detailed in Section \ref{subsec:sel}.

\subsection{Selection}\label{subsec:sel}

In this initial analysis, LArIAT sets out to measure a ``total'' pion cross-section containing multiple processes: elastic and inelastic scattering, charge exchange, pion absorption and pion production. The primary backgrounds to these processes can be classed in two groups: non-pion backgrounds (e.g. interactions from the other constituent particles of the beam), and pions backgrounds (pion decay and capture). The latter group are  
topologically very difficult to distinguish from some signal processes. For this reason, the analysis uses background subtraction to account for pion backgrounds; the selection cuts aim only to remove non-pion backgrounds.

The selection proceeds in three steps:
\begin{enumerate}
 \item A $\mu$/$\pi$/$e$ identification is required from the time of flight detectors (see Figure \ref{fig:tofpid}). \vspace{-2mm}
 \item A clean match is then required between the extrapolated trajectory of the identified particle in the two wire chambers downstream of the magnets and the start of a track in the TPC. \vspace{-2mm}
 \item The identified TPC track is then examined and vetoed if it displays the profile of an electromagnetic shower (to remove electrons from the selected sample. \vspace{-2mm}
\end{enumerate}

This results in a highly pure sample of selected pions, as shown in Table \ref{tab:theonlytable}.

\begin{table}[h!]
\begin{center}
\begin{tabular}{|l|c|c|c|c|c|c|}
\hline
 Particle species                        & $\pi^-$ & $e^-$  & $\gamma$ & $\mu^-$ & $K^-$   & $\bar{p}$ \\
 \hline
 Beam composition before cuts  & 48.4\%  & 40.9\% & 8.5\%    & 2.2\%   & 0.035\% & 0.007\% \\
 \hline
 Percentage passing selection  & 74.5\%  & 3.6\%  & 0.9\%    & 90.0\%  & 70.6\%  & \\
 \hline
\end{tabular}
 \caption{A table of the beam composition in terms of particle species (using the magnets to select negatively-charged particles), and the proportion of particles of each species passing the selection. 
 The proportion of antiprotons passing the selection was deemed too small to merit study.}
 \label{tab:theonlytable}
 \end{center}
\end{table}

\subsection{Result}

\begin{figure}[h!]
 \centerline{\includegraphics[width=0.4\textwidth]{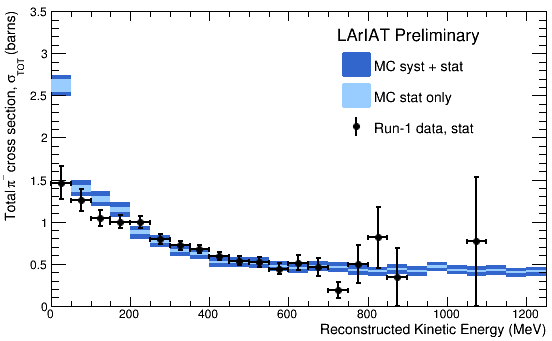}}
 \caption{LArIAT's first measured $\pi^-$-argon cross-section.}
 \label{fig:result}
\end{figure}

Applying the selection to the LArIAT Run 1 negative charge data (approximately three weeks' running in the beam's low-energy tune and five weeks at the high-energy tune) resulted in 2,290 events being selected from an initial sample of 32,064. Analysing 
these events with the method described in Section \ref{subsec:def} gives the extracted cross-section shown in Figure \ref{fig:result}.

This is the world's first measurement of a $\pi^-$ cross-section on argon. The agreement between data and simulation appears good except in the first energy bin, in which deficiencies in the Monte Carlo simulation are suspected. Full simulation of the 
LArIAT beamline is under development, and may resolve this discrepancy. The systematic uncertainties shown at this point are also preliminary, consisting of:

\begin{itemize}
 \item Uncertainty on the $\frac{dE}{dX}$ calibration --- 5\% \vspace{-2mm}
 \item Uncertainty on the energy loss from dead material upstream of the TPC (see Equation \ref{eq:ke}) --- 3\% \vspace{-2mm}
 \item Uncertainty on the contamination from through-going muons --- 3\% \vspace{-2mm}
 \item Uncertainty on the reconstructed momentum from the wire chambers --- 3\% \vspace{-2mm}
\end{itemize}

At this time the uncertainty from pion decay and capture backgrounds remains unaddressed. The assessment of all systematics is still ongoing; as well as improving our analysis techniques, we also expect to make significant gains from bringing in the 
currently unused auxiliary systems (i.e. the aerogel Cherenkov detectors and the muon range stack) to better parameterise the muon backgrounds.

\section{Conclusion}
LArIAT has measured the world's first pion cross-section on argon. This is an important measurement for liquid argon neutrino detectors, and the first of many such measurements LArIAT is positioned to make. Improved analysis techniques, improved 
utilisation of the beamline detectors, and larger amounts of data (incorporating the 5$\frac{1}{2}$ months of data taken in Run 2 in 2016) all offer even more powerful measurements in the near future.


\section*{References}

\begin{thebibliography}{99}
\bibitem{dune} R. Acciarri {\it et al.}, {\it Long-Baseline Neutrino Facility (LBNF) and Deep Underground Neutrino Experiment (DUNE)}, arXiv:1601.02984, 2016
\bibitem{agnt} C. Anderson {\it et al.}, {\it The ArgoNeuT Detector in the NuMI Low-Energy beam line at Fermilab}, \Journal{JINST}{7}{2012}{P10019}.

\end{thebibliography}
\end{document}